\documentclass{elsart}

\usepackage{natbib}

\usepackage{epsfig}

\usepackage{amssymb}

\begin{document}

\begin{frontmatter}




\title{Supernova Remnant 1987A: The Latest Report from the Chandra X-Ray Observatory}


\author[a]{Sangwook Park} 
\ead{park@astro.psu.edu}
\author[b,c]{Svetozar A. Zhekov}
\author[a]{David N. Burrows}
\author[a]{Gordon P. Garmire}
\author[c]{Dick McCray}
\address[a]{Department of Astronomy \& Astrophysics, 525 Davey Lab., 
The Pennsylvania State University, University Park, PA. 16802, USA}
\address[b]{Space Research Institute, Moskovska str. 6, Sofia-1000, Bulgaria}
\address[c]{JILA, University of Colorado, Box 440, Boulder, CO. 80309, USA}

\begin{abstract}
We continue monitoring supernova remnant (SNR) 1987A with the {\it Chandra 
X-ray Observatory}. As of 2004 January, bright X-ray spots in the 
northwest and the southwest are now evident in addition to the bright 
eastern ring. The overall X-ray spectrum, Since 2002 December, can be 
described by a planar shock with an electron temperature of $\sim$2.1 keV. 
The soft X-ray flux is now 8 $\times$ 10$^{-13}$ ergs cm$^{-2}$ s$^{-1}$, 
which is about five times higher than four years ago. This flux increase rate is 
consistent with our prediction based on an exponential density distribution 
along the radius of the SNR between the H{\small II} region and the inner ring. 
We still have no direct evidence of a central point source, and place an 
upper limit of $L_X$ = 1.3 $\times$ 10$^{34}$ ergs s$^{-1}$ on the 3$-$10 keV
band X-ray luminosity.  
\end{abstract}


\end{frontmatter}

\section{Introduction}
\label{intro}
We present the latest results from the {\it Chandra} observations
of supernova remnant (SNR) 1987A. As of 2004 January, we have performed
a total of nine observations of SNR 1987A with the Advanced CCD Imaging
Spectrometer (ACIS) on board {\it Chandra} (Table~1). We have presented
the results from the first seven observations in the literature 
\citep{bur00,park02,mich02,park04}. Previous {\it Chandra}/ACIS 
observations of SNR 1987A have revealed a ring-like X-ray morphology with
an asymmetric X-ray intensity distribution between the eastern and the 
western sides of the SNR. X-ray-bright spots have developed around
the ring, and soft X-ray flux has been nonlinearly increasing. The X-ray 
spectrum was found to be thermal in origin, arising from the
interaction of the blast wave shock with the circumstellar material. 
Previous works \citep{park02,park04} successfully 
demonstrated that these X-ray features, when compared with the optical and radio 
emission features, were consistent with the standard physical model: i.e., 
the SN blast wave shock is approaching the dense inner ring which was
produced by non-spherically symmetric stellar winds from the massive
progenitor \citep{mich00}. In this model, the optically bright spots and 
the soft X-ray spots are produced by the decelerated shock entering the dense 
protrusions on the surface of the inner ring, while the radio and the hard 
X-ray emission features originate from the fast shock propagating into the 
tenuous H{\small II} region between the dense protrusions \citep{park02}. 
Considering its core-collapse, Type II SN origin from a massive progenitor 
\citep{sonn87}, as {\it confirmed} by the detection of the accompanying neutrino 
burst \citep{kosh87}, the presence of a compact stellar remnant is predicted 
for SNR 1987A. There has, however, been no evidence for a pointlike source 
at the center of the remnant, which has been attributed to optically thick stellar
ejecta material surrounding the putative central point source 
\citep{bur00,park02,park04}.

\begin{table}[h]
\begin{center}
Table~1. {\it Chandra} Observations of SNR 1987A\\
\vspace{0.1in}
\label{tbl:tab1}
\begin{tabular}{ccccc}
\hline
ObsID & Date & Instrument & Exposure & Source \\
 & (Age$^{a}$) & & (ks) & Counts\\
\hline
124+1387$^b$ & 1999-10-06 (4609) & ACIS-S + HETG & 116 & 690 \\
122 & 2000-01-17 (4711) & ACIS-S3 & 9 & 607 \\
1967 & 2000-12-07 (5038) & ACIS-S3 & 99 & 9031 \\ 
1044 & 2001-04-25 (5176) & ACIS-S3 & 18 & 1800 \\
2831 & 2001-12-12 (5407) & ACIS-S3 & 49 & 6226 \\
2832 & 2002-05-15 (5561) & ACIS-S3 & 44 & 6429 \\
3829 & 2002-12-31 (5791) & ACIS-S3 & 49 & 9274 \\
3830 & 2003-7-8 (5980) & ACIS-S3 & 45 & 9668 \\
4614 & 2004-1-2 (6157) & ACIS-S3 & 47 & 11856 \\
\hline
\end{tabular}
\end{center}
\vspace{0.1in}
$^a${Day since the SN explosion.}\\
$^b${The first observation was split into two sequences.} 
\end{table}

As the blast wave eventually sweeps through the main body of the dense 
inner ring, a dramatic increase of the X-ray flux and morphological/spectral 
changes of SNR 1987A are expected. We continue X-ray observations of SNR 
1987A with the {\it Chandra}/ACIS in order to monitor the dynamic evolution, 
as well as to discover the central point source, of SNR 1987A. As a 
continuation of the previous works, we here briefly update the morphological 
and spectral evolutions of the X-ray remnant of SNR 1987A with two new 
{\it Chandra} observations.

\section{X-Ray Images}
\label{image}
Fig~1 displays the 0.3$-$8 keV band images of SNR 1987A from the {\it
Chandra}/ACIS observations. SNR 1987A has been significantly brightening 
for the last $\sim$four years. As the blast wave enters the main
body of the dense inner ring, X-ray-bright spots first appeared on the
eastern side of the remnant but have now been developing in the west.
SNR 1987A has thus become closer to a {\it complete} ring than ever.
\begin{figure}[h]
\begin{center}
\hbox{
\psfig{figure=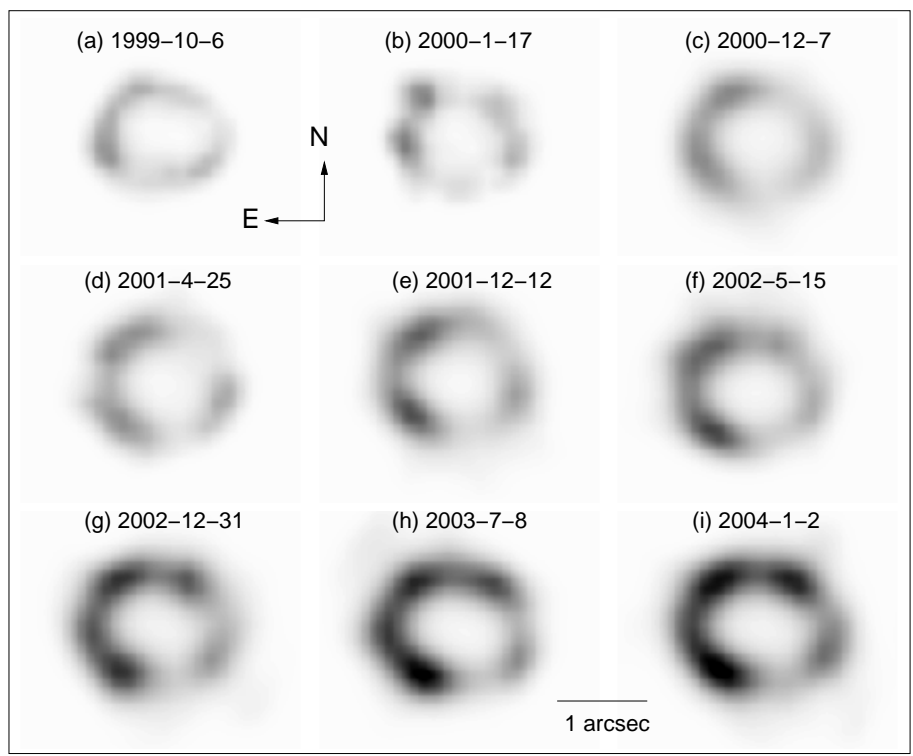,width=0.90\textwidth,angle=0}
}
\vspace{-1.0cm}
\end{center}
\caption{\footnotesize
{Exposure-corrected, broadband (0.3$-$8 keV) X-ray images of SNR 1987A 
from the {\it Chandra}/ACIS observations. Images have been deconvolved 
with the ACIS PSF and then smoothed following the method described in 
literature \citep{bur00,park02}. Darker gray-scales are higher intensities.
}}
\label{fig1}
\end{figure}

\section{X-Ray Spectrum}
\label{spec}
We perform spectral analysis of SNR 1987A following the procedure presented
in Park et al. (2004). The X-ray spectum can be fitted with a single 
temperature plane-parallel shock model \citep{bor01}. The fitted 
foreground column and the elemental abundances are consistent within
statistical uncertainties among the individual spectra. It is also reasonably
expected that the foreground column and the elemental abundances would not
significantly change over a few year-period. We thus fit the observed spectra 
simulaneously, assuming constant N$_H$ and abundances for the last four years,
in order to provide the best constraints on the fitted column and elemental
abundances (Fig~2). We varied elemental abundances for N, O, Ne, Mg, Si, S, 
and Fe, while held in common for all individual spectra. We fixed the abundances 
for He and C at values appropriate for the ring abundances \citep{lund96}, 
and fixed Ca, Ar, and Ni abundances at the values for the Large Magellanic 
Cloud \citep{russell92}, because the contribution of X-ray line emission from 
these elemental species in the fitted energy band (0.4 $-$ 5 keV) is 
insignificant. We varied electron temperature and the ionization timescale
freely for individual spectra in order to trace any significant variations
in these parameters, as perhaps expected from the shock evolution of this
young SNR. For this purpose, we use six spectra with good photon statistics 
($>$6000 counts) for statistically reliable spectral fits. The best-fit 
parameters (with 2$\sigma$ uncertainties) are presented in Table~2 and Table~3.

\begin{figure}[h]
\begin{center}
\psfig{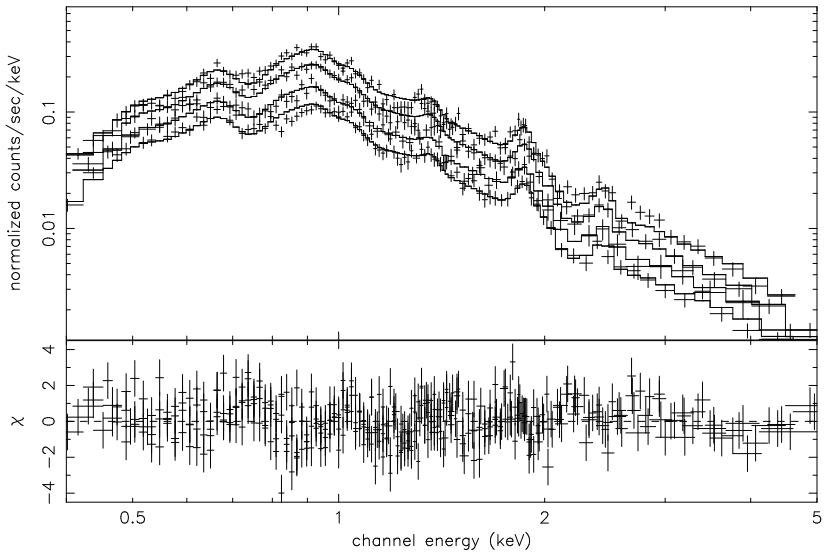}
\vspace{-0.7cm}
\end{center}
\caption{\footnotesize
{Four ACIS spectra of SNR 1987A taken between 2000 December (the lowest flux) 
and 2004 January (the highest flux), separated by $\sim$one year from each other.
The best-fit single-temperature plane-parallel shock model is overlaid with each
spectrum.
}}
\label{fig2}
\end{figure}

Until 2002 December, the electron temperature of SNR 1987A has been decreasing 
while the volume emission measure (EM) was increasing \citep{park04}. The 
latest two observations, however, suggest that the temperature appears to be 
{\it constant} since 2002 December, while the EM was continuously increasing 
(Fig~3). Implications of this recent trend of the electron temperature variation 
are not straightforward. Park et al. (2004) attributed the decrease of electron 
temperature to a physical picture where parts of the shock front enter the 
dense protrusions as the blast wave approaches the inner ring. The latest ACIS 
images, as well as the HST images, indicate that the blast wave shock front is 
now entering the dense circumstellar medium (CSM) {\it all around} the inner 
ring. The simple physical picture with small portions of the shock front 
decelerating as it approaches the dense protrusions, then, might not be adequate 
any longer to describe the overall X-ray spectrum. The X-ray emission might now 
be dominated by the shock entering the main body of the inner ring. The evolution 
of the overall electron temperature in this ``new phase'' of the shock is unclear. 
For instance, it may continue to soften, but not necessarily in the 
previously-reported rate, or it may establish a {\it stable} phase. The details 
of the shock evolution may be complex depending on thermal conditions of the 
plasma (e.g., ionization states and/or electron-ion equilibration behind the 
shock) primarily due to the radial and azimuthal density distribution of the 
inner ring and its environments. We will need upcoming {\it Chandra} observations 
in order to determine whether the latest temperature variation means a truly 
stable phase or only a temporary state in the course of further, continuous 
softening of the overall X-ray spectrum.

\begin{table}[h]
\begin{center}
Table~2. Best-Fit$^a$ Elemental Abundances\\
\vspace{0.1in}
\label{tbl:tab2}
\begin{tabular}{cccccc}
\hline
Element & Abundance$^{b}$ & Element & Abundance$^{b}$ &
Element & Abundance$^{b}$ \\
\hline
 He & 2.57 (fixed$^c$) &  Ne & 0.20$^{+0.02}_{-0.04}$ & Ca & 0.34 (fixed$^d$)  \\
 C  & 0.09 (fixed$^c$) &  Mg & 0.14$^{+0.03}_{-0.02}$ & Ar & 0.54 (fixed$^d$)  \\
 N  & 0.37$^{+0.10}_{-0.08}$ & Si & 0.32$^{+0.05}_{-0.05}$ & Fe & 
0.15$^{+0.01}_{-0.01}$  \\
 O  & 0.09$^{+0.01}_{-0.01}$ & S  & 0.84$^{+0.19}_{-0.18}$ & Ni & 0.62 (fixed$^d$)  \\
\hline
\end{tabular}
\end{center}
\vspace{0.1in}
$^a$ The best-fit N$_H$ = 1.8$^{+0.1}_{-0.2}$ $\times$ 10$^{21}$ cm$^{-2}$.
$\chi$$^2$/$\nu$ = 699.7/569.\\
$^b$Abundances with respect to solar.\\
$^c$ Lundqvist \& Fransson 1996 \\
$^d$ Russell \& Dopita 1992 \\
\end{table}
\begin{table}[h]
\begin{center}
Table~3. Best-Fit Shock Parameters from A Single-Temperature Model$^{a,b}$.\\
\vspace{0.1in}
\label{tbl:tab3}
\begin{tabular}{lccc}
\hline
Date & Electron Temperature & Ionization Timescale & Emission Measure \\
     & (keV) & (10$^{10}$ cm$^{-3}$ s) & (10$^{57}$ cm$^{-3}$) \\
\hline
2000-12-7 & 2.64$^{+0.19}_{-0.20}$ & 3.02$^{+0.50}_{-0.43}$ & 
6.84$^{+0.36}_{-0.27}$ \\
2001-12-12 & 2.48$^{+0.35}_{-0.32}$ & 2.82$^{+0.57}_{-0.43}$ & 
10.20$^{+0.63}_{-0.48}$ \\
2002-5-15  & 2.24$^{+0.29}_{-0.28}$ & 3.23$^{+0.65}_{-0.52}$ & 
12.33$^{+0.78}_{-0.66}$ \\
2002-12-31 & 2.08$^{+0.15}_{-0.14}$ & 3.20$^{+0.48}_{-0.41}$ & 
16.68$^{+0.93}_{-0.75}$ \\
2003-7-8 & 2.11$^{+0.14}_{-0.13}$ & 3.19$^{+0.42}_{-0.38}$ & 
19.14$^{+1.05}_{-0.87}$ \\
2004-1-2 & 2.13$^{+0.13}_{-0.13}$ & 3.29$^{+0.45}_{-0.37}$ & 
22.89$^{+1.23}_{-0.96}$ \\
\hline
\end{tabular}
\end{center}
\vspace{0.1in}
$^a$$\chi$$^2$/$\nu$ = 699.7/569.\\
$^b$ The best-fit N$_H$ = 1.8$^{+0.1}_{-0.2}$ $\times$ 10$^{21}$ cm$^{-2}$.\\
\end{table}
\begin{figure}[h]
\begin{center}
\hbox{
\psfig{figure=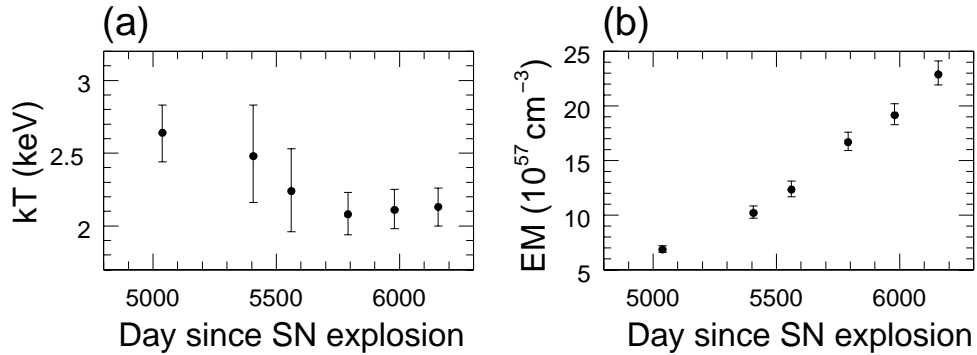,width=\textwidth,angle=0}
}
\vspace{-1.0cm}
\end{center}
\caption{\footnotesize
{The electron temperature and EM variations of SNR 1987A.
}}
\label{fig3}
\end{figure}

Although the overall X-ray spectrum of SNR 1987A can be described by a single 
temperature plane-shock in non-equilibrium ionization (NEI), the physical 
model of the shock-inner ring interaction indicates that X-ray emission most 
likely originates from multiple velocity components of the shock in various
ionization states. Considering such a multi-component shock, a two-temperature 
shock model was implemented in order to fit the observed X-ray spectrum 
\citep{mich02,park04}. The two-temperature shock model may
grossly oversimplify reality, but should be a physically more realistic model
than a single shock. In this model, the observed X-ray spectrum originates 
from two characteristic components. The soft component arises from a slow, 
decelerated shock entering the dense protrusions of the inner ring. The hard
component is produced by a fast shock propagating into the low density 
H{\small II} region.  The slow shock was then found to be in 
collisional ionization equilibrium (CIE), which was reasonable for the shocked 
dense circumstellar material \citep{park04}. Based on these previous results, 
we apply a two-component model to the latest data (Note: As discussed above, 
the overall shock phase for the latest X-ray spectrum might be different from 
that of previous observations. Nonetheless, a two-temperature model should 
still be a useful approximation for the multi-phase shock). As of 2004 
January, an electron temperature of $kT$ $\sim$ 0.23 keV is implied for the 
slow shock in CIE condition, while it is $kT$ $\sim$ 2.2 keV for the fast 
shock in NEI state. The best-fit EMs imply electron densities $n_e$ $\sim$ 6300 
cm$^{-3}$ for the slow shock and $n_e$ $\sim$ 280 cm$^{-3}$ for the fast shock. 
The hard to soft component EM ratio shows a noticeable increase since 2002 
December (Fig~4), which suggests that a significant increase of the density in 
the fast shock might have begun. We note that the volume increase of the hard 
X-ray emitting plasma might also contribute to the observed EM ratio increase. 
We, however, found that the effect from the volume increase is negligible. 
This density increase is consistent with the blast wave, for the most parts, 
now entering the main body of the dense inner ring.
\begin{figure}[h]
\centerline{{\psfig{figure=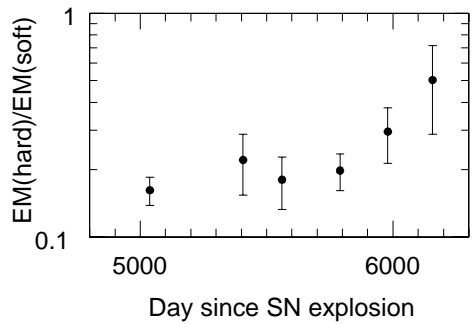,width=0.50\textwidth}}
}
\vspace{-0.15in}
\caption[]{\footnotesize
The hard to soft component EM ratios of SNR 1987A.
}
\label{fig:fig4}
\vspace{-0.1cm}
\end{figure}

\section{X-Ray Lightcurve}
\label{lc}
Fig~5 is the X-ray lightcurve of SNR 1987A. The steep, non-linear increase
of the 0.5$-$2 keV band X-ray flux continues. As of 2004 January, the observed 
0.5$-$2 keV X-ray flux is $f_X$ $\sim$ 8.2 $\times$ 10$^{-13}$ ergs cm$^{-2}$ 
s$^{-1}$ which is $\sim$5 times higher than 1999 October. Following the same 
method described by Park et al. (2004), we fit the lightcurve assuming an 
exponential density distribution along the radius of the inner ring. 
The best-fit model indicates that a density ratio between the high-density inner 
ring and the low-density H{\small II} region ({\it interior} to the inner ring) 
is $\sim$18, and that a scale height of the exponential density distribution
between inner ring and the H{\small II} region is $\sim$0.015 pc at a distance 
of 50 kpc. The estimated density ratio is consistent with the soft to hard
component density ratio derived from the two-component spectral fit.

\begin{figure}[h]
\begin{center}
\hbox{
\psfig{figure=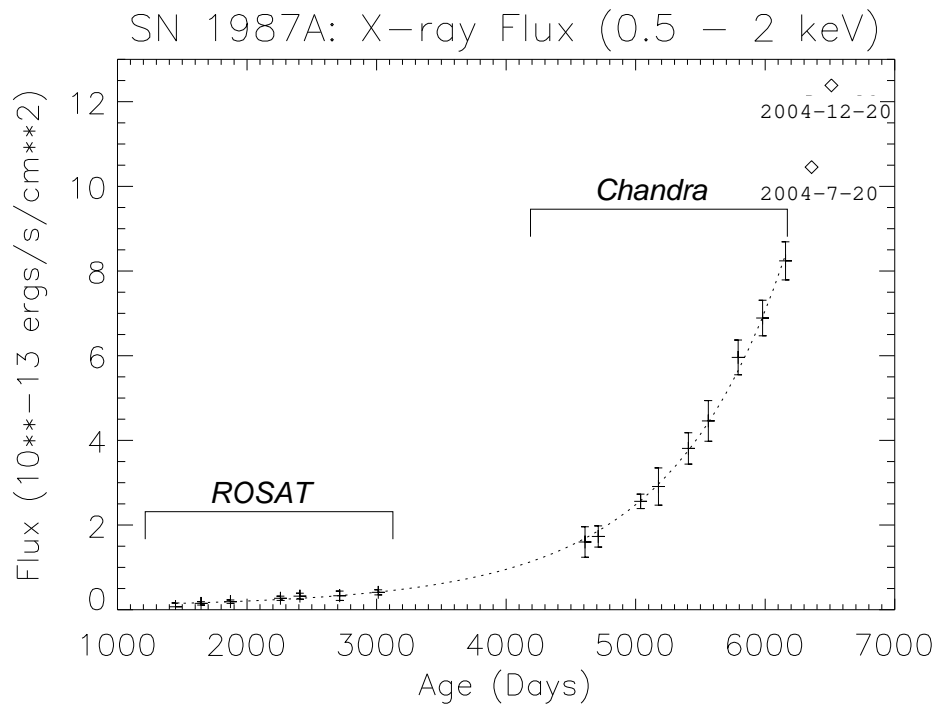,width=\textwidth,angle=0}
}
\vspace{-1.0cm}
\end{center}
\caption{\footnotesize
{The X-ray lightcurve of SNR 1987A. The {\it ROSAT} data points are taken
from Hasinger, Aschenbach, \& Tr\"umper (1996). The dotted curve is the
best-fit model assuming an exponential density profile. Predictions 
for the next two data points are also marked.
}}
\label{fig5}
\end{figure}

\section{Point source}
\label{point}
After $\sim$17 years of the SN explosion, we still find no direct evidence 
of a pointlike source at the center of SNR 1987A. Utilizing a Monte Carlo 
simulation \citep{bur00,park02}, we estimate a 90\% upper limit on the 3$-$8 
keV band point source counts based on the latest ACIS image of SNR 1987A. 
We assume a power law spectrum of a photon index $\Gamma$ = 1.7 for the putative 
neutron star in order to estimate the X-ray flux corresponding to the derived
upper limit on the point source counts. We note that the absorbing column
toward the center of the SNR is highly uncertain depending on the density
structure of the actual stellar ejecta prevailing there. Here we assume, as a
first-order-estimation, a similar absorbing column $N_H$ as measured from the 
spectral fitting of the entire SNR. We then obtain an upper limit of $L_X$ = 
1.3 $\times$ 10$^{34}$ ergs s$^{-1}$ on the 3$-$10 keV band X-ray luminosity 
of the central point source. Considering higher columns, for instance,
if the absorbing column for the point source was an order of magnitude higher 
than that for the entire SNR, this upper limit could be a few times higher
than the above estimate.


\end{document}